\documentclass[a4paper]{article}

\usepackage{INTERSPEECH2022}
\usepackage{graphicx}
\usepackage{amssymb,amsmath,bm}
\usepackage{textcomp}
\usepackage{url}
\usepackage[utf8]{inputenc}
\usepackage{CJKutf8}
\usepackage{multirow}
\usepackage{comment}
\usepackage[colorlinks=true]{hyperref}

\newcommand{\jp}[1]{ \begin{CJK}{UTF8}{ipxm}#1\end{CJK} }

\ninept

\title{ChatGPT-EDSS: Empathetic Dialogue Speech Synthesis Trained from \\ ChatGPT-derived Context Word Embeddings}

\makeatletter
\def\name#1{\gdef\@name{#1\\}}
\makeatother
\name{{\em Yuki Saito$^{1}$, Shinnosuke Takamichi$^1$, Eiji Iimori$^{1}$, Kentaro Tachibana$^2$, and Hiroshi Saruwatari$^1$}}

\address{
$^1$The University of Tokyo, Japan, $^2$LINE Corp., Japan.
}
\email{
yuuki\_saito@ipc.i.u-tokyo.ac.jp
}

\begin{document}
\setlength{\abovedisplayskip}{5pt} 
\setlength{\belowdisplayskip}{5pt} 
\setlength\floatsep{6pt} 
\setlength\intextsep{6pt} 
\setlength\textfloatsep{6pt} 
\setlength{\dbltextfloatsep}{6pt} 
\setlength{\dblfloatsep}{5pt}

\maketitle

\begin{abstract}
We propose \textit{ChatGPT-EDSS}, an empathetic dialogue speech synthesis (EDSS) method using ChatGPT for extracting dialogue context. ChatGPT is a chatbot that can deeply understand the content and purpose of an input prompt and appropriately respond to the user's request. We focus on ChatGPT's reading comprehension and introduce it to EDSS, a task of synthesizing speech that can empathize with the interlocutor's emotion. Our method first gives chat history to ChatGPT and asks it to generate three words representing the intention, emotion, and speaking style for each line in the chat. Then, it trains an EDSS model using the embeddings of ChatGPT-derived context words as the conditioning features. The experimental results demonstrate that our method performs comparably to ones using emotion labels or neural network-derived context embeddings learned from chat histories. The collected ChatGPT-derived context information is available at \href{https://sarulab-speech.github.io/demo_ChatGPT_EDSS/}{our project page.} \\
\noindent{\bf Index Terms}: text-to-speech, empathetic dialogue speech synthesis, dialogue context, ChatGPT, prompt engineering
\end{abstract}

\vspace{-5pt}
\section{Introduction}\label{sect:intro}
\vspace{-3pt}
Dialogue speech synthesis (DSS)~\cite{bruce95}, i.e., text-to-speech (TTS)~\cite{sagisaka88} for spoken dialogue systems, is a crucial technology to actualize natural speech communication between humans and robots. In contrast to TTS, which primarily aims to convey information written in the given input text correctly, DSS requires its speaking style to be more properly controlled in accordance with the dialogue situation (e.g., restaurant reservation~\cite{kim14} and persuasion~\cite{hiraoka16}). Such control is achieved by estimating \textit{context}, e.g., intention~\cite{hojo19} and speakers' emotions~\cite{polzin00}, from dialogue history and conditioning a DSS model by the context~\cite{guo21}.

Empathetic DSS (EDSS)~\cite{saito22studies} is an emerging technology for developing a friendly voice agent that can empathize with an interlocutor. As with empathetic dialogue generation~\cite{rashkin19}, an EDSS model is trained to synthesize speech with an empathetic speaking style using the dialogue context. For instance, Saito et al.'s EDSS method~\cite{saito22studies} uses the speaker's emotion label as the context and improves the quality of synthetic speech. However, this method relies on the annotations of utterance-wise emotion labels for each speaker, which requires the annotators (i.e., \textit{human dialogue advisers}) to deeply understand the empathetic dialogue lines. Although a data-driven context embedding learning~\cite{guo21} can provide a way to control the expressive speaking style of synthetic speech from the chat history, the learned embedding vectors are often hard to interpretable for humans.

\begin{figure}[t]
  \centering
  \includegraphics[width=0.99\linewidth, clip]{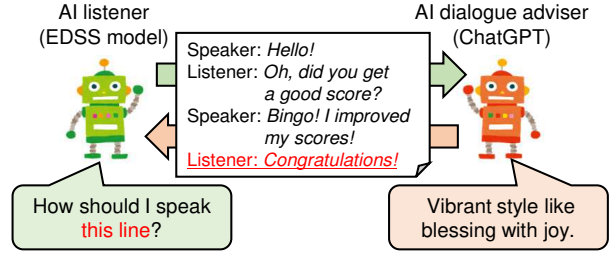}
  \vspace{-8pt}
  \caption{Conceptual dialogue of our ChatGPT-EDSS.}
  \label{fig:concept}
\end{figure}

In text-based dialogue paradigm, \href{https://chat.openai.com/chat}{ChatGPT} (generative pre-trained Transformer), a state-of-the-art artificial intelligence (AI) chatbot, has achieved meaningful breakthroughs in various creative applications, such as writing novels and song lyrics. It is based on GPT-3~\cite{brown20gpt3}, which has been fine-tuned using supervised learning and reinforcement learning for generating response texts preferred by humans~\cite{ouyang22}. This learning mechanism enables ChatGPT to deeply understand the content and purpose of input text prompts and appropriately respond to the user's requests. Although this superior reading comprehension has the potential to extract the dialogue context from the given chat history, the applicability of ChatGPT to spoken dialogue technologies has not yet been investigated. 

To this end, we propose {\it ChatGPT-EDSS}, a ChatGPT-powered EDSS method using ChatGPT as \textit{an AI dialogue adviser}, as shown in Fig.~\ref{fig:concept}. Our method first gives dialogue history to ChatGPT as the prompt and asks it to generate three words related to the context: intention, emotion, and speaking style for each dialogue line. Then, it trains an EDSS model using embedding vectors of the three words as conditional features. We present our methodology to collect ChatGPT-derived context words, training method for ChatGPT-EDSS, and experimental evaluation using a Japanese speech corpus of empathetic dialogue. The contributions of this study are as follows:
    \vspace{-2pt}
    \begin{itemize} \leftskip 1mm \itemsep -0mm
        \item We investigate a way to introduce ChatGPT to spoken dialogue research, especially in EDSS that requires deep understanding of dialogue context to properly control the speaking style of synthetic speech.
        \item We present the prompt design to obtain meaningful context by using ChatGPT and analyze the obtained context words.
        \item We demonstrate that the use of ChatGPT-derived context word embeddings can achieve naturalness and style similarity of synthetic speech comparable to that of ground-truth emotion labels or deeply-learned context embeddings~\cite{guo21}.
    \end{itemize}
    \vspace{-3pt}

\begin{figure*}[t]
  \centering
  \includegraphics[width=0.8\linewidth, clip]{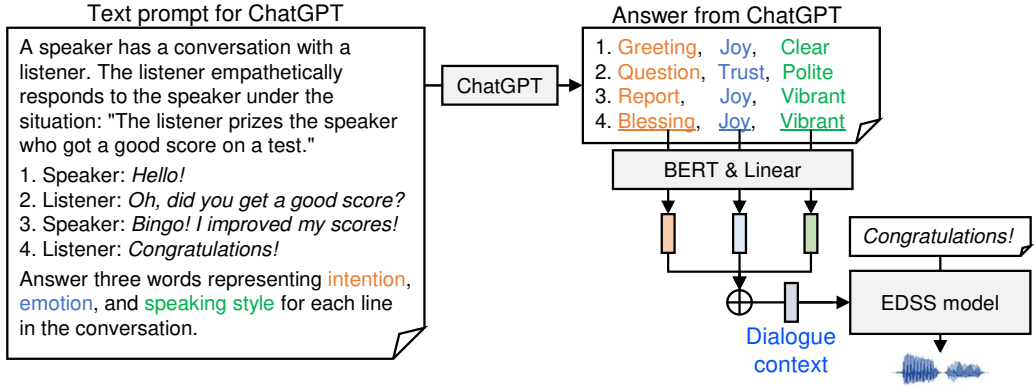}
  \vspace{-10pt}
  \caption{Overview of our ChatGPT-EDSS.}
  \label{fig:chatgpt_edss}
\end{figure*}

\vspace{-5pt}
\section{Related Work}
\vspace{-3pt}
\subsection{EDSS using explicit and implicit dialogue context}
\vspace{-3pt}
Saito et al. proposed a baseline EDSS method~\cite{saito22studies} using ground-truth emotion labels and embedding vectors of chat history as explicit and implicit dialogue context features, respectively. They introduced a conversational context encoder (CCE)~\cite{guo21} that extracts an embedding vector from lines of chat history as the implicit context feature. Their method outperformed a FastSpeech 2 (FS2)-based TTS model~\cite{ren21} regarding the reproducibility of speaking style in EDSS.

\vspace{-3pt}
\subsection{Ability of ChatGPT}
\vspace{-3pt}
As of March 2023, many researchers have been exploring ChatGPT's ability in real-world situations (e.g., education, evaluation~\cite{leiter23chatgptafter2.5month}) and theory of mind~\cite{kosinski23theoryofmindllm}. In addition, some have introduced ChatGPT into the assessment of human's states via texts, e.g., personalities~\cite{rao23chatgptasesspersonalities} and sentiment~\cite{qin23generalpurposenlpsolover}. These work motivate us to use text dialogue contexts obtained by ChatGPT for enhancing DSS technologies.

\vspace{-3pt}
\subsection{Media creation from prompt}
\vspace{-3pt}
With the advancement of deep generative modeling techniques such as denoising diffusion probabilistic models~\cite{ho20ddpm}, media generation from a prompt has been widely studied. DaLL-E~\cite{ramesh21} is one of the first models to generate a realistic image from an input prompt. GPT-3~\cite{brown20gpt3} is an autoregressive large language model that can continue to generate successive sentences from a given initial text as the prompt. AudioGen~\cite{kreuk23audiogen} and MusicLM~\cite{agostinelli23} are generative models for environmental sound and music from text descriptions, respectively. These technologies offer an intuitive way to control the outcomes of media creation by changing the natural language description in the prompt.

Compared with image or text generation, research on TTS control using text prompts is still developing. One primary reason is the difficulty in constructing a sufficiently large dataset that includes many triplets of text to be spoken, speech, and natural language description to explain the speech. Guo et al.~\cite{guo22prompttts} dealt with this difficulty by asking proprietary experts to write prompts that describe the given speech and diversifying the prompts by using SimBERT~\cite{simbert}. Although their dataset contains more than 150,000 data that can be used for training a text-prompt-aware TTS model, such a method for constructing the dataset is very costly and hard to generalize.

\vspace{-5pt}
\section{ChatGPT-EDSS}
\vspace{-3pt}
As shown in Fig.~\ref{fig:chatgpt_edss}, our ChatGPT-EDSS consists of two steps: 1) collection of dialogue context words using ChatGPT and 2) training of an EDSS model using the context words.

\vspace{-3pt}
\subsection{Collection of dialogue context words}\label{subsect:annotation}
\vspace{-3pt}
We ask ChatGPT to generate the dialogue context words from empathetic dialogue lines. As shown in the left part of Figure~\ref{fig:chatgpt_edss}, the text prompt consists of 1) description of dialogue setting, 2) dialogue lines, and 3) request for answering context words.

\textbf{Dialogue setting description} explains the roles of the speaker and listener as well as the dialogue situation to ChatGPT. We empirically found that presenting the dialogue situation improved the relevance of outcomes.

\textbf{Dialogue lines} describe the content of the conversation per turn. The format is a sequence of ``[turn ID] [speaker's name] [content]'' for each dialogue line. We limit the maximum turns in one prompt to 5 because when ChatGPT is asked to answer about long dialogues, it tends to hang in the middle of the answer. If one dialogue consists of more than five turns, we divide it into multiple queries overlapping two turns from the previous query. For example, a dialogue taking 10 turns is divided into prompts for 1--5, 3--7, 5--9, and 7--10 dialogue turns.

\textbf{Request for answering context words} asks ChatGPT to generate words describing the dialogue context for each line. We consider three kinds of context words: 1) dialogue intention~\cite{hojo19}, 2) emotion~\cite{polzin00}, and 3) speaking style. The categories of answers for the emotion and speaking style are \{ neutral, joy, anticipation, anger, disgust, sadness, surprise, fear, trust \} (i.e., neutral and eight emotions defined by Plutchik~\cite{plutchik80}) and \{ cute, cool, quiet, polite, intellectual, honest, clear, gentle, gravelly, vibrant \}, respectively.

\vspace{-3pt}
\subsection{EDSS model training using context word embeddings}
\vspace{-3pt}
We extract embeddings of the collected words by using BERT~\cite{devlin19} and condition an EDSS model by the embeddings. Specifically, we define a ChatGPT-derived dialogue context vector as the sum of the word embeddings and train the EDSS model to predict empathetic dialogue speech from an input text and the context vector. This method regards ChatGPT as an interactive context estimator and replaces the CCE used in Saito et al.'s baseline EDSS method~\cite{saito22studies} with ChatGPT.

\begin{table*}[t]
\centering
\caption{Averaged reliability scores and most frequent words appeared in each emotion category of collected context words}
\label{tab:annotation_results}
\vspace{-3mm}
\begin{tabular}{c|c|c|c|c}
\hline
\hline
& Reliability score & Intention & Emotion & Style \\
\hline
Neutral & 3.95 & \jp{問いかけ} (question) & \jp{期待} (anticipation) & \jp{落ち着いた} (quiet) \\
Happy & 4.04 & \jp{祝福} (blessing) & \jp{喜び} (joy) & \jp{穏やか} (gentle) \\
Angry & 3.66 & \jp{共感} (\textbf{empathy}) & \jp{信頼} (trust) & \jp{丁寧} (polite) \\
Sad & 4.03 & \jp{共感} (\textbf{empathy}) & \jp{悲しみ} (sadness) & \jp{丁寧} (polite) \\
\hline
\hline
\end{tabular}    
\end{table*}

\vspace{-3pt}
\subsection{Discussion}
\vspace{-3pt}
Our ChatGPT-EDSS relates to text-predicted global style tokens (TP-GSTs)~\cite{stanton18}, a TTS method that predicts an expressive speaking style from a text-derived prosody embedding. From this perspective, our method uses ChatGPT to extract style-related words from dialogue lines and predicts the prosody embedding from the extracted words. Although TP-GSTs can improve the quality of synthetic speech better than a Tacotron-based TTS model~\cite{wang17tacotron}, it cannot consider the dialogue history to train the TTS model, which is essential for reproducing an empathetic speaking style in EDSS~\cite{saito22studies}. However, one can introduce a similar idea that uses text-derived prosody embedding to predict the weight for each GST (i.e., predicting combination weights proposed in \cite{stanton18}) in the ChatGPT-EDSS training.

From another perspective, one can regard our ChatGPT-EDSS as weakly supervised learning~\cite{zhou18wsl} of expressive TTS that uses the context words to condition the EDSS model instead of the ground-truth emotion label. We discuss the reliability of the collected context words later in Section~\ref{subsect:word_analysis} because ChatGPT may generate an improper answer from the given prompt.

\vspace{-5pt}
\section{Experimental evaluation}
\vspace{-3pt}
We evaluated our ChatGPT using the STUDIES~\cite{saito22studies} corpus including Japanese empathetic spoken dialogues. The dialogue domain was chit-chat between a female teacher (empathetic listener) and students in a school.

\vspace{-3pt}
\subsection{Experimental conditions}
\vspace{-3pt}
This section describes the conditions for our evaluation.

\textbf{Conditions for context word collection:} We collected the context words for the long (10--20 turns) and short (4 turns) dialogues included in the STUDIES corpus using ChatGPT. The numbers of dialogues were 150 and 720, respectively. We employed 31 workers who asked ChatGPT to generate the context words and annotated the reliability score for each answer with an integer between 1 (``very unreliable'') and 5 (``very reliable''). The workers completed these procedures by 1) accessing Google Sheets prepared by us, 2) copying the text prompts contained in the first field, 3) pasting the prompts to the ChatGPT query field, 4) copying and pasting the ChatGPT answer to the second field, and 5) filling the reliable score in the third field\footnote{We can automate this procedure excluding the reliability scoring  because the ChatGPT API has become accessible since March 2, 2023.}. We asked workers to resend the query to ChatGPT when 1) it failed to generate the context words, 2) the answer included a sentence other than the context word (e.g., the speaker's name or the original dialogue line), and/or 3) the language of the answer was not Japanese (e.g., English or Chinese).

\textbf{Conditions for EDSS:} We trained an EDSS model of the STUDIES teacher with the collected context words. Following Saito et al.'s study~\cite{saito22studies}, we used 726, 72, and 72 dialogues for training, validation, and evaluation, respectively. We downsampled the speech data to 22,050~Hz. We used the validation data to choose hyperparameters for the following models, whose parameters were randomly initialized.

\textbf{Acoustic model for EDSS:} We used FS2~\cite{ren21} as an acoustic model that predicted a mel-spectrogram from text with \href{https://github.com/Wataru-Nakata/FastSpeech2-JSUT}{PyTorch implementation for Japanese TTS}. We followed the settings of a neural network architecture and speech parameter extraction of this implementation. We used the WORLD vocoder~\cite{morise16world,morise16d4c} to estimate $F_0$. The optimizer was Adam~\cite{kingma14adam} with an initial learning rate $\eta$ of 0.0625, $\beta_1$ of 0.9, and $\beta_2$ of 0.98. We first pretrained FS2 using the JSUT corpus~\cite{takamichi20ast}, a Japanese speech corpus including about 10 hours of a female speaker's speech, with 200K iterations. We then fine-tuned it by using the STUDIES training data with 100K iterations.

\textbf{Neural vocoder:} We used a HiFi-GAN vocoder~\cite{kong20} for speech waveform generation from a mel-spectrogram with \href{https://github.com/jik876/hifi-gan}{PyTorch implementation} provided by the first author of the HiFi-GAN paper. We trained HiFi-GAN by using the same training data as that for FS2 with 350K iterations. The optimizer was Adam with $\eta$ of 0.0003, $\beta_1$ of 0.8, and $\beta_2$ of 0.99.

\vspace{-3pt}
\subsection{Analysis of ChatGPT-derived context words}\label{subsect:word_analysis}
\vspace{-3pt}
Table~\ref{tab:annotation_results} lists the results of the context word collection summarized in accordance with the ground-truth emotion labels of the STUDIES teacher. First, we found that the averaged reliability scores were more than 3.6 for all emotion labels. Second, the most frequent intention words for ``Angry'' and ``Sad'' utterances were ``\textbf{empathy}.'' Third, the most frequent emotion word corresponded to the ground-truth emotion label except for ``Neutral'' and ``Angry.'' Finally, the most frequent style words consisted of ``quiet,'' ``gentle,'' and ``polite.'' These results suggest that ChatGPT 1) actually understands the intention of ``empathetic'' dialogue, 2) provides reliable weak labels for expressive TTS  to some extent, and 3) roughly estimates the STUDIES teacher's speaking style as moderate. 

\begin{table}[t]
\centering
\caption{Numbers of unique context words for each ground-truth emotion label}
\label{tab:numbers_of_unique_words}
\vspace{-3mm}
\begin{tabular}{c|r|r|r}
\hline
\hline
& Intention & Emotion & Style \\
\hline
Neutral & 206 & 130 & 42 \\
Happy   & 76 &  35 & 19 \\
Angry   & 17 &  17 & 8 \\
Sad     & 40 & 53 & 19 \\
\hline
\hline
\end{tabular}    
\end{table}

\begin{figure}[t]
  \centering
  \includegraphics[width=0.7\linewidth, clip]{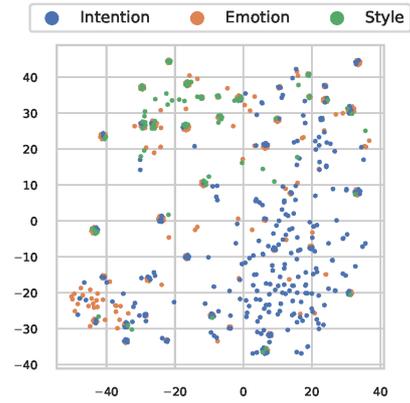}
  \vspace{-10pt}
  \caption{t-SNE plot of BERT embeddings extracted from unique context words}
  \label{fig:emb}
\end{figure}

Table~\ref{tab:numbers_of_unique_words} lists the number of unique context words for each ground-truth emotion label of the STUDIES teacher's utterances. First, the intention words were very diverse and consisted of more than 100 unique words for ``Neutral'' utterances. However, we found that 79\% of these words only appeared five times or less. We observed a similar tendency from the results shown in the third and fourth columns in Table~\ref{tab:numbers_of_unique_words}, despite the fact that we defined the categories of emotion and style in advance. We can confirm these diversities from the t-SNE visualization~\cite{maaten08t_sne} of context word embeddings by BERT shown in Fig.~\ref{fig:emb}, although the different categories tend to form roughly different clusters. These results indicate that ChatGPT 1) can generate various candidate words for describing the context and 2) does not necessarily satisfy the pre-specified requirements for the word generation.

\vspace{-3pt}
\subsection{Subjective evaluations}
\vspace{-3pt}
We conducted subjective evaluations to investigate whether our ChatGPT-EDSS can reproduce the speaking style of empathetic dialogue speech without degrading naturalness.

\textbf{Evaluation setup:} We conditioned the FS2-based EDSS model on the following factors:
\vspace{-2pt}
\begin{itemize}
    \leftskip 1mm \itemsep -0mm
\item {\bf Emo}: Emotion label annotated by the corpus developers
\item {\bf CCE}: Data-driven context embedding extracted from dialogue history~\cite{guo21}
\item {\bf IES (ours)}: Embeddings of intention, emotion, and style words generated from ChatGPT
\end{itemize}
\vspace{-2pt}
The \textbf{CCE} extracted the context embedding from joint vectors of one-hot encoding of speaker identity (3-dim.) and up to four sentence embedding sequences (one current sentence and up to three previous ones) obtained by using BERT (768-dim.) pretrained by using Japanese text data provided at \href{https://huggingface.co/koheiduck/bert-japanese-finetuned-sentiment}{here}. The dimensionality of the context embedding was 256, and we prepared a linear layer to project the BERT-derived word embedding onto the 256-dimensional feature space. 

As explained in Section~\ref{subsect:annotation}, one sentence may have multiple context words due to the overlapping procedure. In that case, we aggregated the embeddings of multiple words for each intention, emotion, and style by simply taking the average of the embeddings. This aggregation may improve the robustness of our ChatGPT-EDSS towards the large variation of context words described in Section~\ref{subsect:word_analysis}.

\textbf{Evaluation criteria:} We conducted a five-scale mean opinion score (MOS) test on the naturalness of synthetic speech. We presented 30 speech samples to listeners in random order. Listeners rated the naturalness of each speech sample from degrees of 1 (``very unnatural'') to 5 (``very natural''). We recruited 50 listeners using our crowdsourcing subjective evaluation system. We also conducted a five-scaled MOS test on speaking-style similarity of synthetic speech. Listeners first listened to the reference speech reconstructed from a natural mel-spectrogram with HiFi-GAN and then scored the presented synthetic speech regarding the similarity of speaking style from degrees of 1 (``very dissimilar'') to 5 (``very similar''). We recruited 50 listeners using our crowdsourcing subjective evaluation system.

\textbf{Evaluation results:} Table~\ref{tab:mos} shows the evaluation results. We found that using \textbf{IES} as the conditional features for the EDSS model performed comparably to using \textbf{Emo} or \textbf{CCE}. This result demonstrates that we can use ChatGPT as the context embedding extractor from empathetic dialogue lines instead of using the emotion label or conventional data-driven context embedding vector. We also observed that the two EDSS models using \textbf{Emo} only and the combination of \textbf{Emo} and \textbf{CCE} slightly degraded the naturalness MOS, while the one using both \textbf{Emo} and \textbf{IES} scored higher MOS values regarding the naturalness and similarity. One reason is the fine-grained (and possibly unlimited) emotion categories represented by the emotion words showing in Table~\ref{tab:numbers_of_unique_words}, which enhances the expression ability of the EDSS model compared with the limited number of emotion categories in the ground-truth label (only four). 

\begin{table}[t]
\centering
\caption{Results of MOS tests on speech naturalness and speaking style similarity with their 95\% confidence intervals}
\label{tab:mos}
\vspace{-3mm}
\begin{tabular}{ccc|rr}
\hline
\hline
\multicolumn{3}{c|}{Method} & \multicolumn{2}{c}{MOS} \\
Emo & CCE & IES \textbf{(ours)} & Naturalness & Similarity \\
\hline
\checkmark & & &            3.43$\pm$0.14 & 3.20$\pm$0.15 \\
& \checkmark & &            3.54$\pm$0.14 & 3.24$\pm$0.14 \\
& & \checkmark &            3.52$\pm$0.14 & 3.19$\pm$0.15 \\
\checkmark & & \checkmark & 3.52$\pm$0.14 & 3.21$\pm$0.14 \\
\checkmark & \checkmark & & 3.43$\pm$0.14 & 3.24$\pm$0.14 \\
& \checkmark & \checkmark & 3.49$\pm$0.14 & 3.20$\pm$0.14 \\
\hline
\hline
\end{tabular}
\end{table}

\begin{figure}[t]
  \centering
  \includegraphics[width=0.9\linewidth, clip]{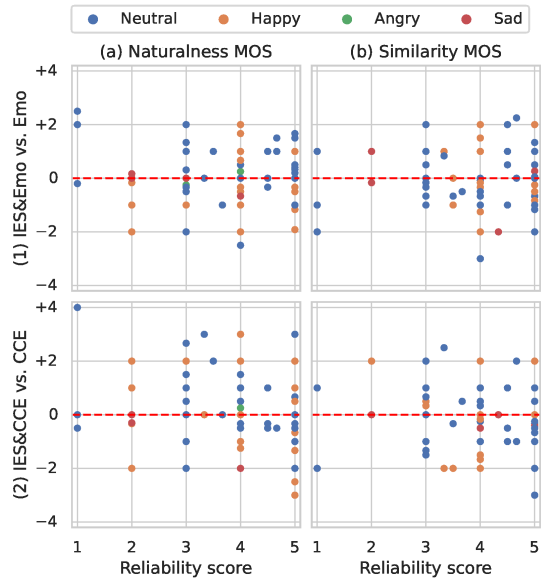}
  \vspace{-10pt}
  \caption{MOS improvement with respect to reliability score of ChatGPT answer}
  \label{fig:rel_to_mosi}
\end{figure}

To further investigate the effects caused by introducing ChatGPT-derived context words in EDSS, we calculated the differences between the naturalness and similarity MOS of 1) \textbf{IES\&Emo} and \textbf{Emo} and 2) \textbf{IES\&CCE} and \textbf{CCE} with respect to the reliability score for each ground-truth emotion label. Figure~\ref{fig:rel_to_mosi} shows the results. From this figure, we observe that there is no correlation between the reliability score and the MOS improvement and that the improvement varies widely even when the reliability score is 5. This result suggests that, although the introduction of \textbf{IES} does not negatively affect the quality of synthetic speech, we still require countermeasures against the large diversity of ChatGPT responses discussed in Section~\ref{subsect:word_analysis}.

\vspace{-5pt}
\section{Conclusion}
\vspace{-3pt}
We proposed ChatGPT-EDSS, a ChatGPT-powered empathetic dialogue speech synthesis (EDSS) method using word embeddings of ChatGPT answers as the dialogue context. Our method first gives text-based chat history to ChatGPT as a prompt and asks it to generate three words related to the dialogue context: intention, emotion, and speaking style for each dialogue line. Then, it trains an EDSS model using embedding vectors of the three words as conditional features. The evaluation results demonstrated that our ChatGPT-EDSS performed comparably to ones using emotion labels or deeply-learned context embeddings extracted from chat histories. Our future work is to investigate the effect of the dialogue domain in ChatGPT-EDSS and to examine whether ChatGPT's hallucination occurs in our method or not.

\textbf{Acknowledgements:}
This research was conducted as joint research between LINE Corporation and Saruwatari-Takamichi Laboratory of The University of Tokyo, Japan.

   \ninept
   \bibliographystyle{IEEEtran}
   \bibliography{template}

\end{document}